\documentclass[preprint,12pt, a4paper]{elsarticle}

\usepackage{amssymb}

\usepackage{listings}
\usepackage{xcolor}
\usepackage{url}
\definecolor{codegreen}{rgb}{0,0.6,0}
\definecolor{codegray}{rgb}{0.5,0.5,0.5}
\definecolor{codepurple}{rgb}{0.58,0,0.82}
\definecolor{backcolour}{rgb}{0.95,0.95,0.95}

\lstdefinestyle{mystyle}{
    backgroundcolor=\color{backcolour},   
    commentstyle=\color{codegreen},
    keywordstyle=\color{magenta},
    numberstyle=\tiny\color{codegray},
    stringstyle=\color{codepurple},
    basicstyle=\ttfamily\footnotesize,
    breakatwhitespace=false,         
    breaklines=true,                 
    captionpos=b,                    
    keepspaces=true,                 
    numbers=left,                    
    numbersep=5pt,                  
    showspaces=false,                
    showstringspaces=false,
    showtabs=false,                  
    tabsize=2
}

\lstMakeShortInline[columns=fixed]|

\usepackage{float}
\restylefloat{table}

\journal{SoftwareX}

\begin{document}

\begin{frontmatter}



\title{\textsc{Pykat}: Python package for modelling precision optical interferometers}


\author{Daniel D. Brown}
\address{OzGrav, University of Adelaide, 5005, Australia}
\author{Philip Jones}
\author{Samuel Rowlinson}
\author{Andreas Freise}
\address{School of Physics and Astronomy, and Institute of Gravitational Wave
    Astronomy,University of Birmingham, Edgbaston, Birmingham B15 2TT, United Kingdom}

\author{Sean Leavey}
\address{Max Planck Institute for Gravitational Physics (Albert Einstein Institute) }
\address{Leibniz Universit\"at Hannover}

\author{Anna C. Green}
\address{University of Florida, Gainesville, FL 32611, USA}

\author{Daniel T\"oyr\"a}
\address{OzGrav, Centre for Gravitatonal Astrophysics, Australian National University,
Acton, The Australian Capital Territory 2601, Australia}

\begin{abstract}
\textsc{Pykat} is a Python package which extends the popular optical interferometer modelling software \textsc{Finesse}. It provides a more modern and efficient user interface for conducting complex numerical simulations, as well as enabling the use of Python's extensive scientific software ecosystem. In this paper we highlight the relationship between \textsc{Pykat} and \textsc{Finesse}, how it is used, and provide an illustrative example of how it has helped to better understand the characteristics of the current generation of gravitational wave interferometers.
\end{abstract}


\end{frontmatter}

\begin{table}[H]
\resizebox{\textwidth}{!}{
\begin{tabular}{ccc}
\hline
\textbf{Nr.} & \textbf{Code metadata description} &  \\
\hline
C1 & Current code version & v1.2 \\
\hline
C2 & Permanent link to code/repository used for this code version & For example: \url{https://git.ligo.org/finesse/pykat} \\
\hline
C4 & Legal Code License   & GPL v2 \\
\hline
C5 & Code versioning system used & git \\
\hline
C6 & Software code languages, tools, and services used & Python \\
\hline
C7 & Compilation requirements, operating environments \& dependencies & 
numpy,
scipy,
six,
h5py,
pandas,
matplotlib,
tabulate,
click
\\
\hline
C8 & If available Link to developer documentation/manual & \url{http://www.gwoptics.org/learn/} \\
\hline
C9 & Support email for questions & Finesse-dev@nikhef.nl \\
\hline
\end{tabular}
}
\end{table}



\section*{Introduction}
\label{}
In order to design the precision interferometers at the heart of advanced gravitational wave detectors, a multitude of bespoke and generic optical simulation software packages have been developed~\cite{sim_book_chapter}. One of the most popular approaches to modelling these interferometers is to use the frequency domain picture, which allows the user to study the quasi-steady state behaviour of an interferometer. Such tools are used to model noise couplings, design control systems, and how defects in an interferometer affect the behaviour of both. As the complexity of the interferometers being designed and constructed as upgrades to the second generation and for third generation facilities has increased, so has the complexity of the modelling packages to keep up with the problems the community has faced. Several packages using this method have been developed: Melody~\cite{melody}, Optickle~\cite{optickle}, MIST~\cite{MIST}, and \textsc{Finesse}~\cite{FINESSE, Freise04}.

The interferometers typically used for gravitational wave detection are well described by a set of linear coupling equations between the various components. This set of equations grows rapidly when the number of optical components, frequencies of the light fields, and higher-order spatial modes of those light fields increases. All of the above packages take in some description of the interferometer, construct a sparse matrix representing the set of equations describing the system, then solve it for various excitations and configurations.

\textsc{Finesse} is one of the most feature-rich packages available and has had extensive usage throughout the gravitational wave community. It is programmed in C, is free, open source, and cross platform. It consists of a single binary executable, called |kat|, that performs the required numerical computations. The user provides a text file, known as a kat-file, that describes both the model of an interferometer and the type of simulation to be executed, using a domain-specific language, kat-script. The binary then outputs a text file with the output data requested by the user. This interaction is performed entirely via a command-line interface from a system terminal.

Despite \textsc{Finesse} being very computationally efficient, the simplistic user interface began to become a hindrance as the complexity of the modelling tasks grew. However, due to the simple command-line interface it was easily called and manipulated via other higher level programming languages such as Bash, Perl, MATLAB, and eventually Python. The Python wrapper, \textsc{Pykat}, is the latest and by far the most advanced wrapper available. It enables a user to easily perform complex sets of modelling tasks efficiently and enables \textsc{Finesse} to be used in conjunction with the wide variety of Python packages, for example, for generating plots, for solving complex optimisation problems or for producing reduced-order models~\cite{romhom}.

\textsc{Pykat} has grown extensively from what it was first conceived to be. Originally it aimed to just simply read kat-script and produce an object-oriented description of the model and simulation to produce a graphical user\--in\-ter\-face (GUI). The object-oriented description however proved much more useful for developing and running simulations. We then developed \textsc{Pykat} to be much more focused on providing better interoperability with \textsc{Finesse} binary. As of today, the majority of all \textsc{Finesse}-based modelling tasks are now undertaken through \textsc{Pykat}.

In this article we highlight how \textsc{Pykat} works with \textsc{Finesse} to run simulations in a more efficient and procedural manner. We also provide an illustrative example of how \textsc{Pykat} was used in practice to better understand the behaviour of real gravitational wave interferometer.

\section{Installation}

\textsc{Pykat} is available from \url{http://www.gwoptics.org/pykat/} with links a git repository with the source code, including the most recent version under development. The release version can be installed through the Conda package management system \url{https://anaconda.org/gwoptics/pykat} or from PyPi \url{https://pypi.org/project/PyKat/}. At the time of writing Conda is the recommended method to install \textsc{Pykat}, because Conda also hosts the \textsc{Finesse} binaries for multiple platforms, and \textsc{Finesse} must be installed for \textsc{Pykat} to work.

\section{Wrapping up \textsc{Finesse} with Python}

\begin{figure*}
	\centering
		\includegraphics[width=\textwidth]{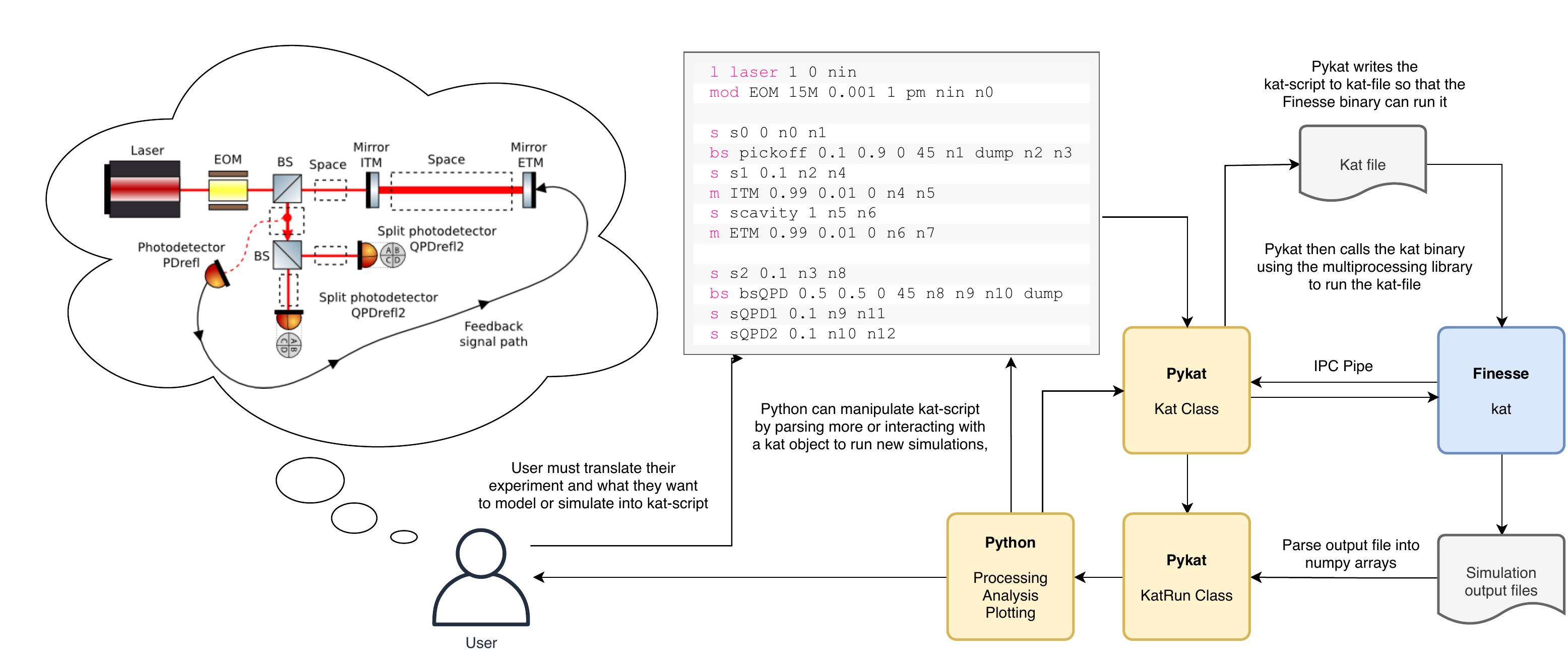}
	\caption{A schematic of the typical workflow a user has with \textsc{Pykat}. Initially the user must have a concrete idea of the physical setup they want to model, and then express this setup in kat-script. This script can then be loaded into a \textsc{Pykat} object which allows the user to run multiple simulations, perform more complex analyses, and generate plots. Behind the scenes, \textsc{Pykat} will seamlessly setup and call the \textsc{Finesse} binary to run the required simulations.
	}
	\label{FIG:pykat_overview}
\end{figure*}

\textsc{Pykat} should be thought of as a wrapper to run \textsc{Finesse} simulations. Thus, to effectively use it the user must be familiar with how \textsc{Finesse} works computationally and physically, what it can and cannot compute, and how to use kat-script. For that, the best reference is the \textsc{Finesse} user manual~\cite{finesse_manual} and the review article by~\citet{livrev}. Additional learning materials can also be found at~\url{http://www.gwoptics.org/learn/}, which contains online notebooks for learning about interferometry as well as the materials for several workshops we have taught using \textsc{Pykat}. 

Kat-script has three distinct classes of instruction: \textit{components}, \textit{detectors}, and \textit{commands}. \textit{Components} are objects such as mirrors, beam splitters, lenses, etc. Components can be connected together to form a network. Each component has a number of nodes to represent the incoming and outgoing light fields; the edges of the network describe how that light goes from one component to the next. \textit{Detectors} are objects that produce an output when a simulation is run. For example, if you want to measure the optical power at a particular location in the interferometer, you would place a `photodiode' detector there. Detectors can be more abstract, however, such as for measuring particular properties of a laser beams shape that would otherwise be difficult or impossible in a real setup. \textit{Commands} are those instructions that configure a component or detector, or setup the type of simulation to run. The Python code for these three classes of instructions can be found in |pykat.components|, |pykat.detectors|, and |pykat.commands|.

When using \textsc{Finesse} in the traditional way, the user would write their kat-script into a text file and then from the command line call the binary to run it, |kat my_simulation.kat|. This would then produce a text file called |my_simulation.out| containing simulation output data and display a plot of the detector outputs as a function of some model parameter being varied.

Figure~\ref{FIG:pykat_overview} shows an overview of how a user would typically interact with \textsc{Pykat} instead. 
\textsc{Pykat} aims to offer Python classes that represent all available kat-script instructions. Models and simulations can be built in \textsc{Pykat} through two interfaces: by \textit{parsing the kat-script}; or by using the object-oriented interface. Kat-script instructions are \textit{parsed} into an instance of \textsc{Pykat}'s |Kat| object. It is constructive for the user to imagine a |Kat| object like a kat-file that you would pass to the \textsc{Finesse} binary at the command-line, except the kat-file is dynamically generated by \textsc{Pykat}.

Parsing the kat-script converts the instructions into Python objects which are added to the |Kat| object which the user can interact with. These objects can be manipulated as required, such as changing properties, parsing further commands, etc. Once the user is ready a simulation can be run. The |Kat| object will first construct a kat-file from the objects contained within it, then pass it to the \textsc{Finesse} binary. 
\textsc{Pykat} and \textsc{Finesse} then communicate with each other during the running of a simulation via an inter-process communication pipe, allowing data and commands to be exchanged. Once completed, the outputted text file with the simulation result is loaded and processed into a |KatRun| object, which provides an easy interface for the user to access all the detector outputs via |numpy| arrays.

Consider the example in Listing~\ref{lst:simple}where we create a model with a laser emitting light that impinges upon a mirror. In this model we add two detectors to measure the complex amplitude of the light fields being reflected and transmitted through the mirror. These commands are parsed into a |Kat| object. We then change the mirror's reflectivity and transmissivity and run it. The output file is then accessed much like a dictionary using the detectors' names. This allows the user to quickly plot and manipulate any outputs.

\begin{lstlisting}[language=Python,
                   label={lst:simple},
                   caption=Simple Pykat usage to run a Finesse simulation]
import pykat
    
kat = pykat.finesse.kat()
kat.parse("""
l l1 1 0 n1  % laser with P=1W
s s1 1 n1 n2 % space of 1m length
m m1 0.5 0.5 0 n2 n3 % mirror 

% an `amplitude' detector for transmitted light
ad ad_t 0 n3      
ad ad_r 0 n2      

% changing the transmittance of the mirror `m1'
xaxis m1 t lin 0.5 0 100 
% plotting amplitude and phase of the results 
yaxis abs:deg     
""")
# Change some properties of the mirror
kat.m1.R = 0.6
kat.m1.T = 0.4

# Generate the kat-file and then run Finesse
out = kat.run()

# Detector outputs are accessed from the
# returned output object by name
print(out['ad_r'])
# Can easily do math with various outputs
# as they are numpy arrays
P_total = abs(out['ad_r'])**2 + abs(out['ad_r'])**2

plt.plot(out.x, P_total)
\end{lstlisting}

The second interface option is to directly use the object-oriented interface to build and run simulations as shown in Listing~\ref{lst:oo}. Most commonly used \textsc{Finesse} features and components are supported by Python objects, however not all of the advanced features are. This interface is ideal for programmatically building a model of an optical experiment as it does not rely on excessive string operations and manipulations. Parameters of components and detectors can be set more easily using keyword-arguments, rather than relying on the strict and terse kat-script instructions.
\begin{lstlisting}[language=Python,
                   label={lst:oo},
                   caption=Object-oriented interface building a cavity and scanning the laser frequency.]
import pykat
from pykat.components import laser, mirror, space
from pykat.detectors import pd
from pykat.commands import xaxis

kat = pykat.finesse.kat()
kat.add( laser('l1', 'n1', P=1) )
kat.add( mirror('m1', 'n1', 'n2', R=0.99, T=0.01) )
kat.add( space('s1', 'n2', 'n3', L=4000) )
kat.add( mirror('m2', 'n3', 'n4', R=0.99, T=0.01) )
kat.add( pd('P_t', 0, 'n4') )
kat.add( xaxis('lin', (0, 100e3), kat.l1.f, 100) )
out = kat.run()
\end{lstlisting}
Although this interface is likely more favoured by programmers, it does remove the more direct connection the user has to the \textsc{Finesse} binary. Thus, they must be mindful of what \textsc{Finesse} is actually running. For more complex modelling tasks, and for debugging, this information is sometimes required explicitly. The user can quickly see the current kat-script that would be sent to \textsc{Finesse} if it was run by simplying printing the |kat| object, |print(kat)|. However, it is entirely possible to use a mix of both interfaces.

While the example in Listing~\ref{lst:simple} shows a task that could in principle be accomplished using only \textsc{Finesse} and a single kat-script, users often require running multiple simulations to produce specific results, which would be a tedious manual task without \textsc{Pykat}. A common example of such a task is the optimisation of some experimental parameter. Using \textsc{Pykat} we can easily access a variety of packages for such a task, such as |Scipy|. Listing~\ref{lst:base_models} depicts a toy optimisation problem where we wish to find the optimal transmissivity of the second mirror in the cavity to optimise the amount of power being transmitted. When run, the result should be that it is equal to the transmission of the first mirror~\cite{livrev}, in what is known as an \textit{impedance matched cavity} configuration.
\begin{lstlisting}[language=Python,
                   label={lst:base_models},
                   caption=Toy example for optimising transmitted power. Highlights the base-model-deepcopy pattern for running multiple simulations.]
import pykat
import scipy

base = pykat.finesse.kat()
base.parse("""
l l1 1 0 n1  
s s1 1 n1 n2 
m m1 0.99 0.01 0 n2 n3
attr m1 Rc -10
s s2 1 n3 n4 
m m2 1 0 0 n4 n5
attr m2 Rc  10
cav c m1 n3 m2 n4

pd P_t n5   
pd P_r n2

maxtem 0
""")

base.verbose = False
base.noxaxis = True

def func(T):
    if T > 1 or T <= 0: return float('inf')
    kat = base.deepcopy()
    kat.m2.R = 1-T
    kat.m2.T = T
    out = kat.run()
    return -1 * out['P_t']

res = scipy.optimize.minimize(func, x0=0.1)
# Update the base model now with
# the new optimised parameters
base.m2.setRTL(R=1-res.x, T=res.x)
\end{lstlisting}
Listing~\ref{lst:base_models} also introduces an important coding pattern in \textsc{Pykat}: \textit{deepcopying a Kat object}. Here we borrow from the Python term of deep-copying, whereby we make an entirely new \verb+Kat+ object that is completely separate from the original. Again this can be thought of as making a separate kat-file that can be altered to run an alternative simulation. The pattern is that the user should first create a \textit{base} model that describes the optical setup and detectors. For each simulation that is required, a new deepcopy of base model should be made, extra simulation code added to it, then the new model can be run.

Although many attempts have been made, \textsc{Finesse} does not significantly benefit from parallelisation when running a single simulation. Experience has shown, it is more optimal to run many instances of \textsc{Finesse} simultaneously. \textsc{Pykat} offers an easy to use interface for quickly running multiples simulations. The interface uses the |Ipyparallel| package, which allows it to utilise anything from the multiple cores on a single machine to distributing the tasks to an entire cluster. Example~\ref{lst:parakat} provides an overview of how this works.
\begin{lstlisting}[language=Python,
                   label={lst:parakat},
                   caption=Code for running parallel
                   \textsc{Finesse} simulations.]
import pykat
from pykat.parallel import parakat

pk = parakat()

kat = pykat.finesse.kat()
kat.parse("""
...
[kat-script instructions]
...
""")

for value in values:
    kat.component.parameter = value
    pk.run(kat)
    
outs = pk.getResults()
\end{lstlisting}

\section{Illustrative example: debugging noise couplings in Advanced LIGO}

A typical problem encountered by instrument scientists working on gravitational wave detectors is to understand how various sources of noise in the interferometer couple into the output channel that measures passing gravitational waves. 
Sometimes these couplings can be explained or explored with simple analytical models. However, as the complexity of the instrument has grown, simple models often no longer represent what is happening in reality. This is where \textsc{Finesse} becomes particularly useful as it allows us to simulate such complex setups. 

\begin{figure}
	\centering
		\includegraphics[width=0.5\textwidth]{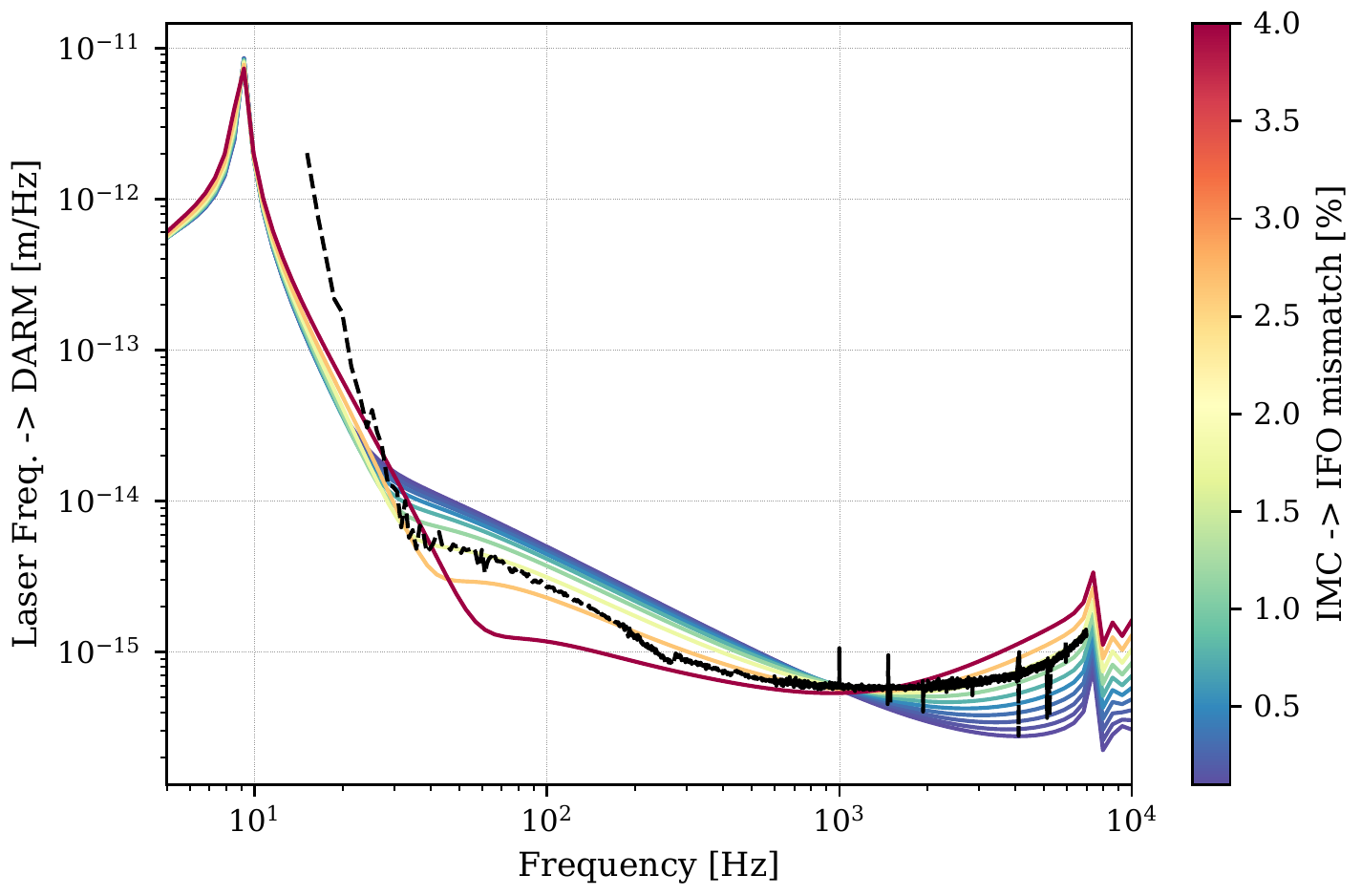}
\caption{Shown is a measured transfer function showing how frequency noise couples into the gravitational wave channel at LIGO as a function of frequency (black, dashed). Also shown is the model configuration found to have the best fit and how a variation in the shape of the input laser beam changes this noise coupling, reproducing a similar shape to the measurement around 100~Hz and at 1--8~kHz.}
	\label{FIG:plot_freq_tf}
\end{figure}

In this example, we highlight a modelling task undertaken on measurements from one of the Advanced LIGO~\cite{ALIGO2015} interferometers. The issue at hand was understanding how the frequency noise of the input laser couples into the gravitational wave output channel. The measured transfer function for this noise coupling is shown in Figure~\ref{FIG:plot_freq_tf}. Simple models predict that the frequency noise coupling drops as $\propto 1/f$, however the measurement shows some structure around 100~Hz and then smoothly rises above 1~kHz.

To model this type of problem we first have to produce a realistic model of the Advanced LIGO detectors. This has to include realistic defects that we know or expect to be present. Then we model different specific configurations and output how the frequency noise injected at the laser source propagates to the gravitational readout channel. Finally we can compare the result to the measured values and ask which configuration best fits the data.

A \textsc{Finesse} model that describes a realistic interferometer in a  gravitational wave detector can easily reach 300+ instructions. When using a single kat-file and \textsc{Finesse}, such a model can become large and cumbersome to work with.
\textsc{Pykat} has trivialised creating models of the advanced detectors, for example here we show how we can create an Advanced LIGO model and initialise it to be at its correct operating point in just a handful of lines:
\begin{lstlisting}[language=Python,
                    label={lst:aligo},
                   caption=Getting an Advanced LIGO model]
import pykat
from pykat.ifo import aligo
import pykat.ifo.aligo.plot
# Create a new base model of the as designed
# Advanced LIGO interferometer
base = aligo.make_kat()
# This then sets up all the locks, error signals,
# and operating points for the interferometer model
aligo.setup(base)
# Finally plot the quantum noise limited strain
# sensitivity
aligo.plot.strain_sensitivity(base)
# Plot all the length sensing error signals
aligo.plot.error_signals(base)
\end{lstlisting}
This base model represents an ideal interferometer, no defects are included. The \verb+setup()+ function is optimising all the mirror positions to ensure they are at the correct \textit{operating point}~\cite{livrev}. Simply put, this means the mirrors are held at the correct microscopic position to ensure that the optical fields resonate in the interferometer as designed.

From this idealised model of LIGO we can begin to introduce defects: incorrectly sized laser beams; deviations in mirror properties, such as optical loss, radii of curvature, or alignment; and errors in how well we microscopically position the mirrors. All these effects are likely to be present in the experiment, we have ranges of how strong each of these deviations might be but the exact amount and relative proportions of each is not well known.

In this case, given the number of free parameters, we chose to run a simple Monte-Carlo search computing the frequency noise coupling for each. Using the \textsc{Pykat} parallelisation features the parameter space was quickly explored using a computer cluster.
For each of these potential defected configurations we have to make sure our interferometer still operates correctly; otherwise we may model the noise coupling in an interferometer state that cannot be used for gravitational wave detection. One important figure of merit is ensuring we can still effectively sense and control all the lengths of the interferometer, also known as \textit{locking the interferometer}. \textsc{Pykat} includes functionality (See |kat.IFO.lock_drag()|) for performing a technique called \textit{lock-dragging}~\cite{ddb-thesis} which allows a user to find a suitable operating point easily---which when not performed correctly, or at all, results in unphysical and incorrect models and is a common stumbling point for new users.

Of the order 100k simulations were run looking at different geometric configurations of the interferometer and input laser beam states. The configuration with the best-fit is shown in Figure~\ref{FIG:plot_freq_tf}. It was found that generating a noise coupling with this shape requires that some astigmatism must be present in the surface figure of the LIGO test mass optics, while the input laser shape is incorrect by several percent. This configuration might not be the exact state of the real interferometer, however it highlights features that can be  investigated further experimentally, to either rule out or confirm the hypothetical source of this extra noise coupling. Clearly, performing this analysis using only \textsc{Finesse} would have been impractical.

\section{Impact}

In the past 20 years \textsc{Finesse} alone has had a significant impact in the field of gravitational wave science~\cite{FinesseImpact}. However, with \textsc{Pykat} we have significantly increased the capabilities and reach of our simulation software.
First of all, \textsc{Pykat} has enabled significantly more complex modelling tasks to be undertaken with \textsc{Finesse} with ease. It achieves this by providing a more modern and adaptable user interface allowing \textsc{Finesse} to be connected with a wide variety of scientific packages in Python. 
Expert users can now code up their knowledge into functions, such as \verb+aligo.setup()+ shown in Listing~\ref{lst:aligo}, that other users can easily call on. It also significantly reduces errors from working with multiple kat-files, by providing the tools to do so in a more procedural manner.

The use of \textsc{Pykat} from Python scripts has enabled researchers to share their modelling via version controlled repositories with ease leading to increased research software sustainability~\cite{10.1109/CHASE.2019.00039}. In particular, Jupyter Notebooks~\cite{jupyter-notebook} have provided a far more interactive format for conducting, documenting, and distributing numerical modelling tasks and results---which has led them to becoming the preferred method of using \textsc{Pykat} and \textsc{Finesse} now for many users.

\textsc{Pykat} was the key tool to leverage the entire Python ecosystem and to benefit from the rising popularity of the Python language in the science community. More importantly, the notebook format has enabled us to produce more engaging and interactive learning materials, not only for using \textsc{Finesse}, but also for teaching undergraduate students and early career researchers about precision interferometry in a more hands-on environment. This has resulted in an unprecedented amount of training in interferometer simulation for gravitational wave detection, through online resources or workshops. This is of particular importance for young researchers in new gravitational wave group. This year we created dedicated training material and organized a 'hackathon' as part of the effort to train students for the new LIGO detector planned in India~\cite{LIGOIndia}. \textsc{Pykat} is the key component around which such training and teaching is being developed. Multiple examples of similar workshops and the respective materials can be found at~\url{http://www.gwoptics.org/learn}. 

\section{Conclusion}

In this paper we have described the interferometer simulation software \textsc{Pykat} and the symbiotic relationship between \textsc{Pykat} and \textsc{Finesse}. \textsc{Pykat} provides a new modern and efficient Python interface that enables us to simulate precision optical experiments previously not possible. We outlines an illustrative example of how it enabled us to perform a complex modelling task to better understand current noise couplings in of one of the LIGO gravitational wave interferometers. The success of \textsc{Pykat} has enabled us to provide better learning materials for students, improved the software sustainability, and allows researchers to tackle more complex problems.

Although \textsc{Pykat} is primarily for modelling optical experiments using \textsc{Finesse}, it includes several other features that have not been discussed here. It has also become the home for several other computational tools which users may find helpful, such as, a Fast-fourier-transform optical modelling tool, based on the software package \textsc{Oscar}~\cite{degallaix2008oscar}, ABCD Gaussian beam propagation code, routines for computing higher-order-mode Gaussian beam scattering, and generating reduced-order-models for light scattering~\cite{romhom}.

\section{Conflict of Interest}
The authors confirm that there are no known conflicts of interest associated with this publication and there has been no significant financial support for this work that could have influenced its outcome.

\section*{Acknowledgements}

The authors would like to extend their thanks to the international gravitational wave community for their feedback and support in developing both \textsc{Pykat} and \textsc{Finesse}. DDB and DT were supported by the ARC grant CE170100004. SL has been supported by the Deutsche Forschungsgemeinschaft (DFG, German Research Foundation) under Germany's Excellence Strategy - EXC-2123 QuantumFrontiers - 390837967. AF has been supported by the Science and Technology Facilities Council (STFC) and by a Royal Society Wolfson Fellowship which is jointly funded by the Royal Society and the Wolfson Foundation. The authors would like to thank Aaron Jones for providing the 'Birmingham Environment for Software Testing' (BEST) which we used for testing \textsc{Pykat} during development. DDB also thanks Craig Cahillane for the fruitful discussions on modelling interferometers and the data for the frequency noise coupling. The Authors would also like to thank the LIGO-Virgo Collaboration for use of the computing cluster for running our \textsc{Finesse} models. This document has been given the LIGO DCC number P2000104.  The authors have no competing or financial interests to declare.

\bibliographystyle{unsrt}

\bibliography{pykat}

\begin{thebibliography}{10}

\bibitem{sim_book_chapter}
David Reitze, Peter Saulson, and Hartmut Grote.
\newblock {\em Advanced Interferometric Gravitational-Wave Detectors}.
\newblock World Scientific, 2019.

\bibitem{melody}
Raymond~G. Beausoleil.
\newblock {Melody}.
\newblock \url{https://dcc.ligo.org/public/0034/G010301/000/G010301-00.pdf},
  2001.

\bibitem{optickle}
Matt Evans.
\newblock {Optickle simulation software}.
\newblock \url{https://git.ligo.org/IFOsim/Optickle2}, 2004.

\bibitem{MIST}
Gabriele Vajente.
\newblock Fast modal simulation of paraxial optical systems: the mist open
  source toolbox.
\newblock {\em Classical and Quantum Gravity}, 30(7):075014, 2013.

\bibitem{FINESSE}
Daniel~David Brown and Andreas Freise.
\newblock Finesse.
\newblock \url{http://www.gwoptics.org/finesse}, May 2014.
\newblock {You can download the binaries and source code at
  \url{http://www.gwoptics.org/finesse}.}

\bibitem{Freise04}
A~Freise, G~Heinzel, H~L\"{u}ck, R~Schilling, B~Willke, and K~Danzmann.
\newblock Frequency-domain interferometer simulation with higher-order spatial
  modes.
\newblock {\em Classical and Quantum Gravity}, 21(5):S1067--S1074, 2004.
\newblock Finesse is available at \url{http://www.gwoptics.org/finesse}.

\bibitem{romhom}
D~Brown, R~J~E Smith, and A~Freise.
\newblock Fast simulation of gaussian-mode scattering for precision
  interferometry.
\newblock {\em Journal of Optics}, 18(2):025604, jan 2016.

\bibitem{finesse_manual}
Andreas {Freise}, Daniel {Brown}, and Charlotte {Bond}.
\newblock {Finesse, Frequency domain INterferomEter Simulation SoftwarE}.
\newblock {\em arXiv e-prints}, page arXiv:1306.2973, June 2013.

\bibitem{livrev}
Charlotte Bond, Daniel Brown, Andreas Freise, and Kenneth~A. Strain.
\newblock Interferometer techniques for gravitational-wave detection.
\newblock {\em Living Reviews in Relativity}, 19(1):3, 2017.

\bibitem{ALIGO2015}
{LIGO Scientific Collaboration}.
\newblock Advanced {LIGO}.
\newblock {\em Classical and Quantum Gravity}, 32(7):074001, mar 2015.

\bibitem{ddb-thesis}
Daniel~David Brown.
\newblock {\em Interactions of light and mirrors : advanced techniques for
  modelling future gravitational wave detectors}.
\newblock PhD thesis, 2016.

\bibitem{FinesseImpact}
Andreas Freise and Daniel~David Brown.
\newblock {Finesse, Hostory and Impact}.
\newblock \url{http://www.gwoptics.org/finesse/impact.php}, 2017.

\bibitem{10.1109/CHASE.2019.00039}
M\'{a}rio~Rosado de~Souza, Robert Haines, Markel Vigo, and Caroline Jay.
\newblock What makes research software sustainable? an interview study with
  research software engineers.
\newblock In {\em Proceedings of the 12th International Workshop on Cooperative
  and Human Aspects of Software Engineering}, CHASE ’19, page 135–138. IEEE
  Press, 2019.

\bibitem{jupyter-notebook}
Thomas Kluyver, Benjamin Ragan-Kelley, Fernando P{\'e}rez, Brian Granger,
  Matthias Bussonnier, Jonathan Frederic, Kyle Kelley, Jessica Hamrick, Jason
  Grout, Sylvain Corlay, Paul Ivanov, Dami{\'a}n Avila, Safia Abdalla, and
  Carol Willing.
\newblock Jupyter notebooks -- a publishing format for reproducible
  computational workflows.
\newblock In F.~Loizides and B.~Schmidt, editors, {\em Positioning and Power in
  Academic Publishing: Players, Agents and Agendas}, pages 87 -- 90. IOS Press,
  2016.

\bibitem{LIGOIndia}
Bala Iyer, Tarun Souradeep, CS~Unnikrishnan, Sanjeev Dhurandhar, Sendhil Raja,
  and Anand Sengupta.
\newblock Ligo-india, proposal of the consortium for indian initiative in
  gravitational-wave observations (indigo), 2015.
\newblock \url{https://dcc.ligo.org/LIGO-M1100296/public}.

\bibitem{degallaix2008oscar}
J~Degallaix.
\newblock Oscar: a matlab based fft code.
\newblock {\em Matlab Central File Exchange}, 2008.

\end{thebibliography}

\end{document}